\documentclass[12pt]{article}
\usepackage{amsmath}
\usepackage{amssymb}
\usepackage{epsfig}
\usepackage{euscript}
\def\mathswitchr#1{\relax\ifmmode{\mathrm{#1}}\else$\mathrm{#1}$\fi}

\newcommand {\pslash}{\hbox{$\not\hbox{\kern-2.3pt $p$}$}}

\usepackage{cite}

\usepackage{epic}

\def\alf1{ {\alpha\over\pi} }

\begin{document}
 
\title{IR-Improved Operator Product Expansions in non-Abelian Gauge Theory}
\author{B.F.L. Ward%
    \thanks{Work supported in part by D.o.E. grant DE-FG02-09ER41600.}\\
      Baylor University\\
       \texttt{bfl\_ward@baylor.edu}}
\date{BU-HEPP-12-01,\\
     Apr., 2012}

\maketitle
\begin{abstract}
We present a formulation of the operator product expansion that is infrared finite to all orders in the attendant massless non-Abelian gauge theory coupling constant, which we will oftentimes associate with the QCD theory, the theory that we actually have
as our primary objective in view of the operation of the LHC at CERN. We make contact in this way with the recently introduced IR-improved DGLAP-CS theory and point-out phenomenological implications 
accordingly, with an eye toward the precision QCD theory for LHC physics.
\end{abstract}

%

 
\def\Kmax{K_{\rm max}}\def\ieps{{i\epsilon}}\def\rQCD{{\rm QCD}}

\section{\bf Introduction}\label{intro}\par
With the start-up of the LHC the era of precision QCD, 
by which we mean
predictions for QCD processes at the total precision tag of $1\%$ or better,
is upon us and the need for exact, amplitude-based 
resummation of large higher order effects
is becoming more and more acute. Methods to facilitate the realization
of such resummation are then of particular interest. In this paper, we revisit 
the pioneering use of operator product expansion (OPE) methods, as presented 
by Wilson~\cite{kgw} for short-distance limits of physical processes 
and as applied by 
Gross, Wilczek and Politzer in the QCD~\cite{qcd} theory, especially as it is
realized in the DGLAP-CS~\cite{dglap,cs} theory, 
from the standpoint of 
resummation of
its large infrared effects with an eye toward the attendant application 
of the corresponding parton model representation to LHC precision physics.
In this way, we make contact as well with the recently introduced IR-improved
DGLAP-CS theory in Refs.~\cite{irdglap1,irdglap2,herwiri}.\par
Specifically, it is well-known~\cite{djg-fw,ghdp,hdp1} that the usual formulation of the Wilson expansion in massless gauge theory is infrared divergent: the easiest way to realize this is to note that, already at one-loop, the respective leading twist operator matrix elements between fundamental particle
states are in general infrared divergent and must be evaluated at 
off-shell (Euclidean)
points in massless gauge theory -- see for example Refs.~\cite{djg-fw,ghdp,hdp1}. The result is that the coefficient functions of the Wilson operators
in the OPE which encode the leading
$Q^2$ dependence of the expansion are in general infrared divergent order-by-order in renormalized perturbation theory\footnote{In Ref.~\cite{kgw}, Wilson pointed-out already that the coefficient functions in his expansion could be calculated order by order in perturbation theory and that the n-th order term could contain logarithms of $z^2m^2$ where $z$ is the space-time interval
of the respective two operators and $m$ is the free field mass, so that these logs would be divergent at $m=0$. He also noted that an arbitrary subtraction constant $a$ could be introduced to
convert the argument of these logarithms to $z^2a^2$. This is equivalent to what we have stated in the text with $a=\mu$ where $\mu$ is identified as a Euclidean point in an appropriate convention.}. Of course, all such infrared divergences cancel in physically observable (hadronic) matrix elements of the expansion so that, from the standpoint of such observables, the issue is one of choosing the best rearrangement of the large
infrared effects that remain after all infrared divergences have canceled.
Here, we will resum these large infrared effects. As a result, in what follows, we reformulate the OPE in such away that the respective expansion
components are infrared finite.
As a further result, we show how the new IR-improved DGLAP-CS theory in Ref.~\cite{irdglap1,irdglap2,herwiri} arises naturally in this context. We argue that the IR-improved expansion should be closer to experiment for a given {\it exact} order in the loop expansion for the coefficient functions and respective operator matrix elements. \par
The paper is organized as follows. In the next section, we recapitulate the 
formulation of the OPE following the arguments of Wilson as used in
Refs.~\cite{djg-fw,ghdp,hdp1} for the analysis of deep-inelastic lepton-nucleon scattering~\cite{taylor}, the proto-typical physical application of the method. In Section 3, we show how to improve it so that its hard coefficient functions are IR finite. We also make contact with the new IR-improved DGLAP-CS theory~\cite{irdglap1,irdglap2,herwiri}. In Section 3, we also
sum up with an eye toward phenomenological implications.\par

\section{Review of the OPE}
\begin{figure}[h]
\begin{center}
\epsfig{file=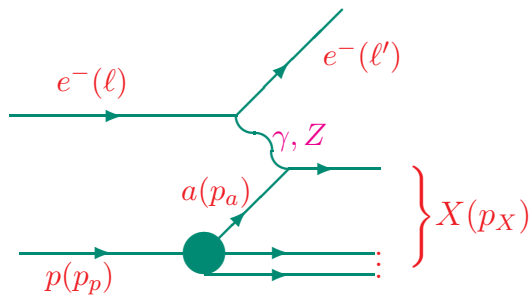,width=70mm}
\end{center}
\caption{Deep inelastic electron-proton scattering: 
$q=\ell -\ell', \; \nu=qp_p/m_p, \; x\equiv x_{Bj}=-q^2/(2m_p\nu)$, $\ell(\ell')$ 
is the four-momentum of the initial(final) $e^-$, $p_A$ is the four-momentum of $A,\; A=a,p$, where $a$ is a parton.}
\label{fig1}
\end{figure}

For pedagogical reasons, we follow the historical development and use the deep inelastic electron-proton scattering problem discussed so effectively by
Bjorken~\cite{bj1} as our starting point: $e^-(\ell)+p(p_p)\rightarrow e^-(\ell')+X(p_X)$. Indeed, his discussion set the framework for the issues we address here. The kinematics and notation are summarized in the Fig.~\ref{fig1}, so that we use $x\equiv x_{Bj}=Q^2/(2m_p\nu)$ for Bjorken's scaling variable which has the interpretation in the attendant parton model as the struck parton's momentum fraction when $\nu=qp_p/m_p$ with $q=\ell-\ell', \; Q^2=-q^2$. In the Fig.~\ref{fig1}, the parton momenta are $p_i(p'_i)$ before(after)
the hard interaction process. 
The limit of Bjorken is then of interest here, in which we take $Q^2 \rightarrow \infty$ with $x$ fixed. In this limit, where here we will for reasons of pedagogy focus on the photon exchange in Fig.~\ref{fig1}~\footnote{As it is well-known, adding in the effects of the Z exchange is straightforward and does not
require any essentially new methods that are not already exhibited by what we do for the photon exchange case.}, the standard methods can be used to represent the imaginary part of the attendant current-proton forward scattering amplitude as 
\begin{equation}
\begin{split}
W^{EM}_{\alpha\beta}(p_p,q)&=\frac{1}{2\pi}\int d^4ye^{iqy}<p|[J^{EM}_\beta(y),J^{EM}_\alpha(0)]|p>\\
&=\qquad (-g_{\alpha\beta}+q_\alpha q_\beta/q^2)W_1(\nu,q^2)\\
&\qquad +\frac{1}{m_p^2}(p_p-qqp_p/q^2)_\alpha(p_p-qqp_p/q^2)_\beta W_2(\nu,q^2)
\end{split}
\label{eq1-ope}
\end{equation}
Here, $J^{EM}_\alpha(y)$ is the hadronic electromagnetic current 
and $W_{1,2}$ are the usual deep inelastic  
the structure functions, which
first were shown to exhibit Bjorken scaling by the SLAC-MIT experiments~\cite{taylor} already at $Q^2\cong 1_+ \text{GeV}^2$, precocious scaling -- we return to this point below. For our
purposes here, henceforward we drop the superscript on $J^{EM}$ so that
$J^{EM}\equiv J$ for ease of notation and we always understand
the average over the spin of the proton even when we do not
indicate so explicitly.
In Bjorken's limit, we have
\begin{equation}
\begin{split}
\lim_{Bj} m_pW_1(\nu,q^2)&=F_1(x)\\
\lim_{Bj} \nu W_2(\nu,q^2)&=F_2(x)
\end{split}
\label{eq2-ope}
\end{equation}
where the scaling limits $F_{1,2}$ only depend on Bjorken's variable $x$ 
and we denote $$\lim_{Bj} \equiv \lim_{Q^2\rightarrow\infty}|_{x-\text{fixed}}.$$ The QCD theory of Gross, Wilczek and Politzer~\cite{qcd} provides a
quantum field theoretic explanation of the observed Bjorken scaling behavior
via Wilson's OPE.\par
Specifically, in Bjorken's limit, the phase in integral over space-time in (\ref{eq1-ope}) oscillates rapidly except in regions where it is bounded so that the value of the integral is dominated by the latter regions, which are well-known
to correspond to the tip of the light-cone~\cite{djg-sbt}, the short-distance
regime. Using Wilson's expansion in this regime, we get the OPE~\cite{djg-fw,ghdp,hdp1} {\small
\begin{equation}
\begin{split}
J_\beta(y)J_\alpha(0)&=\frac{1}{2}g_{\beta\alpha}\left(\frac{\partial}{\partial y}\right)^2\frac{1}{y^2-i\epsilon y_0}{\sum_{n=0}^{\infty}} \sum_{j} C^{(n)}_{j,1}(y^2 - i\epsilon y_0) O^j_{\mu_1 \cdots \mu_n} (0)y^{\mu_1} \cdots y^{\mu_n}\\
   & + \frac{1}{y^2-i\epsilon y_0}{\sum_{n=0}^{\infty}} \sum_{j} C^{(n)}_{j,2}(y^2 - i\epsilon y_0) O^j_{\beta\alpha\mu_1 \cdots \mu_n} (0)y^{\mu_1} \cdots y^{\mu_n}+\cdots,
\end{split}
\label{eq3-ope}
\end{equation}}
where we have neglected gradient terms without loss of content for our purposes here and as usual $\epsilon \downarrow 0$. We also note that $\{ O^j_{\mu_1 \cdots \mu_n} (y)\}$ are traceless, symmetric spin $n$ operators of dimension $n+2$
or of twist = dimension -spin = 2~\cite{djg-sbt}. The $\cdots$ represent operators of higher twist that are suppressed by powers of $q^2$ to any finite order in perturbation theory. The coefficient c-number functions $\{ C^{(n)}_{j,k}\}$ are dimensionless and can be computed in renormalized perturbation theory.\par
Continuing our recapitulation of the methods in Refs.~\cite{djg-fw,ghdp,hdp1},
if we define the spin averaged proton matrix elements of the operators
$ O^j$ via 
\begin{equation}
<p| O^j_{\mu_1 \cdots \mu_n} (0)|p>|_{\text{spin averaged}}=i^n\frac{1}{m_p}{p_p}_{\mu_1} \cdots {p_p}_{\mu_n} M^n_j+\cdots,
\label{eq4-ope}
\end{equation}
where the second $\cdots$ denotes trace-terms, we get the following relationship~\cite{djg-fw,ghdp,hdp1,chm} between the moments of the structure functions and the Fourier
transforms of the coefficient functions:
\begin{equation}
\begin{split}
\int_0^1dx x^n F_1(x,q^2)&=\sum_j \bar{C}^{(n+1)}_{j,1}(q^2)M^{n+1}_j, \\
\int_0^1dx x^n F_2(x,q^2)&=\sum_j \bar{C}^{(n)}_{j,2}(q^2)M^{n+2}_j,
\end{split}
\label{eq5-ope}
\end{equation}
where~\cite{djg-fw}
\begin{equation}
\bar{C}^{(n)}_{j,k}(q^2)=\frac{1}{2}i(q^2)^{n+1}\left(-\frac{\partial}{\partial q^2}\right)^n \int d^4y e^{iqy}{\small \frac{C^{(n)}_{j,k}(y^2)}{y^2-i\epsilon y_0}} .
\label{eq6-ope}
\end{equation} 
The $q^2$ dependence of the $\bar{C}^{(n)}$ is controlled by the Callan-Symanzik equation~\cite{cs} which reads
\begin{equation}
\left[\left(\mu\frac{\partial}{\partial\mu}+\beta(g)\frac{\partial}{\partial g}\right)\delta_{ij}-\gamma^{(n)}_{ij}(g)\right]C^{(n)}_{j,k}=0
\label{eq7-ope}
\end{equation}
where $\mu$ denotes the renormalization scale,
$$\beta(g)=\mu\frac{\partial g}{\partial\mu}$$
for the attendant renormalized coupling $g$,  and the anomalous dimension matrix
$\gamma^{(n)}_{ij}(g)$ is given as 
\begin{equation}
\gamma^{(n)}_{ij}(g)=\left(Z_O^{-1} \mu\frac{\partial }{\partial\mu}Z_O\right)_{ij}|_{g_{0}, \text{regularization fixed}}
\label{eq8-ope}
\end{equation}
where the operators $O^{(n)}_j\equiv O^j_{\mu_1\cdots\mu_n}$ are renormalized via
\begin{equation}
O^{(n)}_i\equiv O^{(n)}_{i,R}=\sum_j O^{(n)}_{j,bare}\left(Z_O^{-1}\right)_{ji}
\label{eq9-ope}
\end{equation}
so that they mix under renormalization in the well-known way~\cite{djg-fw,ghdp}
and we use a standard notation of the renormalized, $R$, and unrenormalized
, bare, operator representatives. $g_0$ is the bare coupling constant.
It is well-known that the solution of (\ref{eq7-ope}) leads to the conclusion that the asymptotic Bjorken limit is controlled by the operators with the smallest eigenvalue for their anomalous dimension matrix in an asymptotically free theory such as the 
QCD~\cite{qcd} which we have in mind here. The implied behavior
for the RHS of (\ref{eq5-ope}) is in agreement with experiment~\cite{taylor}. Here, we want to focus on the IR-improvement of the $C^{(n)}_{j,k},\; M^{n}_j$.
\par
\section{IR-Improved OPE}
The isolation of the infrared aspects of the $C^{(n)}_{j,k}$ is immediate if
we use the fundamental particles in the respective Lagrangian quantum field theory, quarks and gluons in the case of QCD, to evaluate the essential anomalous
dimension matrix elements $\gamma^{(n)}_{ij}(g)$, as this is equivalent to studying deep inelastic scattering from these fundamental particles and takes us
immediately, at least conceptually, to the parton model perspective
studied famously by many~\cite{rpf-part,bj-ps,drlyn,ellis-pol-mach,fur-pet1,fur-pet2}.\par
Specifically, we then focus on the parton level version of hadronic tensor
$W_{\alpha\beta}$ which for definiteness we associate with a fermion $F$
in the underlying asymptotically free theory(QCD):
\begin{equation}
\begin{split}
W^{F}_{\alpha\beta}(p_F,q)&=\frac{1}{2\pi}\int d^4ye^{iqy}<p_F|[J_\beta(y),J_\alpha(0)]|p_F>\\
                         &= (2\pi)^3\sum_{X}\delta(q+p_F-p_X)<p_F|J_\beta(0)|p_X><p_X|J_\alpha(0)|p_F>,
\end{split}
\label{eq10-ope}
\end{equation}
where we use the fact that $q^0>0$ to drop the remaining term in the commutator
and we always average over the spin of the fermion $F$, as we do for the proton$p$. We see clearly from (\ref{eq1-ope}) that the RHS of (\ref{eq10-ope})
and that of (\ref{eq1-ope}) involve the same OPE.\par
We first focus on the matrix element 
\begin{equation}
{\cal M}_{X,\alpha}\equiv <p_X|J_\alpha(0)|p_F>.
\label{eq11-ope}
\end{equation}  
Following Refs.~\cite{irdglap1,irdglap2}, we isolate the dominant virtual IR divergences associated to the incoming line via the formula
\begin{equation}
{\cal M}_{X,\alpha} = e^{\alpha_s\; B_{QCD}} <p_X|J_\alpha(0)|p_F>_{IRI-virt}, 
\label{eq11a-ope}
\end{equation} 
where the virtual infrared function $B_{QCD}$ is given in Refs.~\cite{irdglap1,irdglap2}.
The RHS of this last equation is valid to all orders in $\alpha_s\equiv g^2/(4\pi)$ so that one
computes $<p_X|J_\alpha(0)|p_F>_{IRI-virt}$ from  $<p_X|J_\alpha(0)|p_F>$ by comparing the coefficients of the powers of $\alpha_s$ on both sides of 
(\ref{eq11a-ope}) iteratively.\par
Introducing this result into (\ref{eq10-ope}), we arrive at{\small
\begin{equation}
\begin{split}
W^{F}_{\alpha\beta}(p_F,q)
                         &= (2\pi)^3\sum_{X}\delta(q+p_F-p_X)e^{2\alpha_s \Re B_{QCD}}\;\;{_{IRI-virt}\!<p_F|J_\beta(0)|p_X>}\\
                         & \qquad\qquad\;\;\;\; <p_X|J_\alpha(0)|p_F>_{IRI-virt}.
\end{split}
\label{eq12-ope}
\end{equation}}
We next isolate the leading soft, spin independent real emission infrared function
associated to the incoming line as follows. We first separate $\{X\}$
into its multiple gluon subspaces via
\begin{equation}
\{X\} = \{X: X = X'\otimes \{G_1\otimes\ldots\otimes G_n\}, \text{for some}\; n\ge 0, X'\; \text{is non-gluonic}\}.
\label{eq13-ope}
\end{equation}
Then we have {\small
\begin{equation}
\begin{split}
&e^{2\alpha_s \Re B_{QCD}}{_{IRI-virt}\!<p_F|J_\beta(0)|p_X>}<p_X|J_\alpha(0)|p_F>_{IRI-virt}\\
&=e^{2\alpha_s \Re B_{QCD}}\Big{[}\tilde S_{QCD}(k_1)\cdots \tilde S_{QCD}(k_n)\;\;{_{IRI-virt}<p_F|J_\beta(0)|p_{X'}>}\\
&<p_{X'}|J_\alpha(0)|p_F>_{IRI-virt}+\cdots+ {_{IRI-virt\&real}\!<p_F|J_\beta(0)|p_{X'},k_1,\cdots,k_n>}\\
&<p_{X'},k_1,\cdots,k_n|J_\alpha(0)|p_F>_{IRI-virt\&real}\Big{]},
\end{split}
\label{eq14-ope}
\end{equation}}
\noindent
where the real infrared function $\tilde S_{QCD}(k)$ is given in Refs.~\cite{irdglap1,irdglap2}. 
The IR-improved quantities
$$ {_{IRI-virt\&real}\!<p_F|J_\beta(0)|p_{X}>}<p_{X}|J_\alpha(0)|p_F>_{IRI-virt\&real}$$
are defined iteratively from (\ref{eq11a-ope}),(\ref{eq14-ope}) to all orders in $\alpha_s$ and they no longer contain the infrared singularities from the initial line associated to $B_{QCD}$ and to $\tilde S_{QCD}$, although, because of the non-Abelian infrared algebra of the theory, they do contain other IR singularities which of course cancel in the structure functions by the KNL theorem for massless fundamental fermions. For massive fundamental fermions, these latter singularities also cancel provided we resum the theory as we are doing here accordingly -- see Refs.~\cite{bflw-nbn}.\par 
Introducing the representation in (\ref{eq14-ope}) into  
(\ref{eq10-ope}) we get 
\begin{equation}
\begin{split}
W^{F}_{\beta\alpha}(p_F,q)
                        & = (2\pi)^3\sum_{X}\delta(q+p_F-p_X)e^{2\alpha_s \Re B_{QCD}}\Big{[}\tilde S_{QCD}(k_1)\cdots \tilde S_{QCD}(k_n)\\
  & \qquad\qquad{_{IRI-virt}<p_F|J_\beta(0)|p_{X'}>}<p_{X'}|J_\alpha(0)|p_F>_{IRI-virt}+\cdots\\
&\qquad\qquad\;\;+ {_{IRI-virt\&real}\!<p_F|J_\beta(0)|p_{X'},k_1,\cdots,k_n>}\\
&\qquad\qquad\;\;<p_{X'},k_1,\cdots,k_n|J_\alpha(0)|p_F>_{IRI-virt\&real} \Big{]}\\
           & = \frac{1}{2\pi}\int d^4 y\sum_{X'}\sum_n\frac{1}{n!}\int\Pi_{j=1}^n \frac{d^3k_j}{k^0_j}e^{SUM_{IR}(QCD)}e^{iy(q+p_F-p_{X'}-\sum_j k_j)+D_{QCD}}\\
&\qquad\qquad\;\; {_{IRI-virt\&real}<p_F|J_\beta(0)|p_{X'},k_1,\cdots,k_n>}\\
&\qquad\qquad\;\;<p_{X'},k_1,\cdots,k_n|J_\alpha(0)|p_F>_{IRI-virt\&real}\\
          & = \frac{1}{2\pi}\int d^4y e^{iqy}e^{SUM_{IR}(QCD)+D_{QCD}}\\
& \qquad\qquad\;\;{_{IRI-virt\&real}\!<p_F|}[J_\beta(y),J_\alpha(0)]|p_F>_{IRI-virt\&real},
\end{split}
\label{eq15-ope}
\end{equation}
where we have defined 
\[ {\rm SUM_{IR}(QCD)}=2\alpha_s \Re B_{QCD}+2\alpha_s\tilde B_{QCD}(\Kmax),\]
\[ 2\alpha_s\tilde B_{QCD}(\Kmax)=\int {d^3k\over k^0}\tilde S_\rQCD(k)
\theta(\Kmax-k),\]
 \begin{equation} D_\rQCD=\int{d^3k\over k}\tilde S_\rQCD(k)
\left[e^{-iy\cdot k}-\theta(\Kmax-k)\right],\label{eq15a-ope}\end{equation}
and we stress that (\ref{eq15-ope}) does not depend on $K_{max}$.
Using the standard partonic view, by which we have
\begin{equation}
W_{\beta\alpha}= \sum_a \int_0^1\frac{dx}{x} {\cal F}_a(x)W^a_{\beta\alpha}
\label{eq16-ope}
\end{equation}
for appropriately defined parton distribution functions $\{{\cal F}_a\}$, we introduce the OPE in (\ref{eq3-ope}) into (\ref{eq15-ope}) and use
(\ref{eq1-ope}) to get the IR-improved results
\begin{equation}
\begin{split}
\int_0^1dx x^n F_1(x,q^2)&=\sum_j \tilde{\bar{C}}^{(n+1)}_{j,1}(q^2)\tilde{M}^{n+1}_j, \\
\int_0^1dx x^n F_2(x,q^2)&=\sum_j \tilde{\bar{C}}^{(n)}_{j,2}(q^2)\tilde{M}^{n+2}_j,
\end{split}
\label{eq17-ope}
\end{equation}
where~\cite{djg-fw}
\begin{equation}
\tilde{\bar{C}}^{(n)}_{j,k}(q^2)=\frac{1}{2}i(q^2)^{n+1}\left(-\frac{\partial}{\partial q^2}\right)^n \int d^4y e^{iqy}e^{SUM_{IR}(QCD)+D_{QCD}}{\small \frac{\tilde{C}^{(n)}_{j,k}(y^2)}{y^2-i\epsilon y_0}} 
\label{eq18-ope}
\end{equation}
and now
\begin{equation}
\begin{split}
<p| \tilde{O}^j_{\mu_1 \cdots \mu_n} (0)|p>|_{\text{spin averaged}}&\equiv\\
_{IRI-virt\&real}\!<p| O^j_{\mu_1 \cdots \mu_n} (0)|p>_{IRI-virt\&real}|_{\text{spin averaged}}&=i^n\frac{1}{m_p}{p_p}_{\mu_1} \cdots {p_p}_{\mu_n}\tilde{M}^n_j \\
& \qquad +\cdots,
\end{split}
\label{eq19-ope}
\end{equation}
where the second $\cdots$ again denotes trace-terms and the $\{\tilde{C}^{(n)}_{j,k}\}$ are the respective
(new) IR-improved
OPE coefficient functions.  
The $q^2$ dependence of the $\tilde{\bar{C}}^{(n)}$ is also controlled by the Callan-Symanzik equation~\cite{cs} which now reads
\begin{equation}
\left[\left(\mu\frac{\partial}{\partial\mu}+\beta(g)\frac{\partial}{\partial g}\right)\delta_{ij}-\tilde{\gamma}^{(n)}_{ij}(g)\right]\tilde{\bar{C}}^{(n)}_{j,k}=0
\label{eq20-ope}
\end{equation}
where now the new matrix $\tilde{\gamma}^{(n)}_{ij}(g)$ is determined by the
renormalization properties of the IR-improved matrix elements in (\ref{eq19-ope}) as we will discuss presently. We need to stress that in writing (\ref{eq18-ope}) we work to one-loop order in the various coefficients in this paper.\par
A convenient starting point for obtaining the new matrix $\tilde{\gamma}^{(n)}_{ij}(g)$ is presented by the pioneering analysis
of the authors in Ref.~\cite{fur-pet1,fur-pet2}. Working directly from the the representation in (\ref{eq16-ope}), the authors in Ref.~\cite{fur-pet1}
make contact with the matrix $\gamma^{(n)}_{ij}(g)$ for the unimproved OPE as follows. Focusing for definiteness for the moment on the 
non-singlet operator~\cite{djg-fw} $^N\!O^{F,b}(y)=\frac{1}{2}i^{N-1}S\bar{\psi}(y)\gamma_{\mu_1}\nabla_{\mu_2}\cdots\nabla_{\mu_N}\lambda^b\psi(y)-\text{trace terms}$, where $\nabla_\mu=\partial_\mu+ig\tau^aA^a_\mu$ is the covariant derivative, $\lambda^b$ is a flavor group generator and $S$ denotes symmetrization with respect to the indices $\mu_1\cdots\mu_n$, we have the matrix element between fundamental fermion states, where spin averaging is understood here as well, as
\begin{equation}
<p|^N\!O^{F,b}(y)|p>=\; ^{F,b}\!O^{N}(\alpha_s,\epsilon)p_{\mu_1}\cdots p_{\mu_N}-\text{trace terms}
\label{eq21-ope}
\end{equation} 
where we use $d=4-\epsilon$ dimensions for regularization and the notation  $^{F,b}\!O^{N}(\alpha_s,\epsilon)\equiv M^N_{F,b}$
to make immediate contact with the arguments in Ref.~\cite{fur-pet1}. The renormalized 
matrix element $^{F,b}\!O^{N}(\alpha_s,\epsilon)$ is related to the bare one as we have indicated in (\ref{eq9-ope}):
\begin{equation}
^{F,b}\!O^{N}(\alpha_s,\epsilon,p^2/\mu^2)=Z^{-1}_O(\alpha_s,\frac{1}{\epsilon})\; ^{F,b}\!O^{N}_{bare}((\alpha_s)(\mu^2/p^2)^\epsilon,\epsilon)
\label{eq21a-ope}
\end{equation}
so that collinear divergences are regularized by taking $p^2\ne 0$ in the approach in Ref.~\cite{fur-pet1}.
Using an arbitrary vector $\Delta$ with $\Delta^2=0$ we get
\begin{equation}
^{F,b}\!O^{N}(\alpha_s,\epsilon)=<p|^N\!O^{F,b}_{\mu_1\cdots\mu_N}(y)|p>\Delta^{\mu_1}\cdots \Delta^{\mu_N}/(\Delta p)^N.
\label{eq22-ope}
\end{equation} 
The pole part of $^{F,b}\!O^{N}$ which is the renormalization part $Z^{-1}_O$ can be determined in any gauge by gauge invariance.
We set $\Delta=n$ where $x_{Bj}=np/np_p,\; x=kn/np$ and $nA^a=0$ so that we are in a light-like gauge.
This allows us to write, following Ref.~\cite{fur-pet1}, 
\begin{equation}
^{F,b}\!O^{N}(\alpha_s,\epsilon)=\int_{-1}^{1} dx x^{N-1}\; {^{F,b}\!O(x,\alpha_s,\epsilon)}
\label{eq23-ope}
\end{equation} 
where 
\begin{equation}
 {^{F,b}\!O(x,\alpha_s,\epsilon)}=Z_F[\delta(x-1)+ x\frac{\int d^d k}{(2\pi)^d}\delta(x-\frac{kn}{pn})[\frac{\not\!n}{4kn}T(p,k)\not\!p]
\label{eq24-ope}
\end{equation}
where we use the notation of Ref.~\cite{fur-pet1} so that $T(p,k)$ is the respective fully connected four-point function
and $[\not\!bB$ denotes $\sum_{\alpha\alpha'}b_{\alpha\alpha'}B^{\alpha\alpha'}_{\beta\beta'}$ with
corresponding notation for $B\not\!b]$. $Z_F$ is the fermion field renormalization constant as usual.
By first analytically continuing the LHS of (\ref{eq21a-ope}) 
to $d=4+\epsilon$ 
dimensions with $\epsilon>0$ the authors in Ref.~\cite{fur-pet1} note that
the limit $p^2\rightarrow 0$ gives the RHS as just $Z^{-1}_O(\alpha_s,\epsilon)$ for an appropriate normalization of $\lambda^b$
. The RHS of (\ref{eq23-ope}) may then related to the moments of the densities
of partons in a quark, $\Gamma_S(x,\alpha_s,1/\epsilon)$ in the notation of Ref.~\cite{fur-pet1}, by writing a dispersion relation for $[(\not\!n/(4kn))T(p,k)\not\!p]$ and performing the attendant $k^2$ integral by closing the contour around the dispersive poles(see Sect. 4.2 of Ref.~\cite{fur-pet1}), analytically continuing to $d=4+\epsilon$ dimensions with again $\epsilon>0$ and finally taking the limit $p^2\rightarrow 0$ to get
\begin{equation}
Z^{-1}_O(\alpha_s,\frac{1}{\epsilon})=\int_{-1}^{1}dxx^{N-1}[\Gamma_{qq}(x,\alpha_s,\frac{1}{\epsilon})\theta(x)-\Gamma_{q\bar{q}}(-x,\alpha_s,\frac{1}{\epsilon})\theta(-x)]
\label{eq25-ope}
\end{equation}
where $\Gamma_{qq}(\Gamma_{q\bar{q}})$ is the respective parton density
for a quark(anti-quark) in a quark. The coefficients of $\frac{1}{\epsilon}$
on both sides of this last equation then give the fundamental result, derived in Ref.~\cite{fur-pet1},
\begin{equation}
\begin{split}
-\gamma^{(N)}(\alpha_s)&=2\int_{-1}^{1}dx x^{N-1}[P_{qq}(x,\alpha_s)\theta(x)-P_{q\bar{q}}(-x,\alpha_s)\theta(-x)]\\
      &=2[ P_{qq}(N,\alpha_s)+(-1)^NP_{q\bar{q}}(N,\alpha_s)]
\end{split}
\label{eq26-ope}
\end{equation}
where we define $$F(N)=\int_0^1 dx x^{N-1}F(x)$$ and the $P_{BA}$ are the usual DGLAP-CS~\cite{dglap,cs} splitting kernels defined in the convention of Ref.~\cite{fur-pet1}
and $\gamma^{(N)}(\alpha_s)$ is the respective anomalous dimension of the operator $^N\!O^{F,b}$.\par
To apply this calculation to our new anomalous dimension matrix we IR-improve it at each step as we have shown above (and as we
have shown for the IR-improved DGLAP-CS theory in Refs.~\cite{irdglap1,irdglap2}), so that
we replace $$<p|^N\!O^{F,b}(y)|p>\rightarrow <p|^N\!\tilde{O}^{F,b}(y)|p>$$
as defined in (\ref{eq19-ope}) with the corresponding substitution of 
$^{F,b}\!O^{N}(\alpha_s,\epsilon)$ by the analogous $^{F,b}\!\tilde{O}^{N}(\alpha_s,\epsilon)$. This leads to the relationship 
\begin{equation}
^{F,b}\!\tilde{O}^{N}(\alpha_s,\epsilon,p^2/\mu^2)=Z^{-1}_{\tilde{O}}(\alpha_s,\frac{1}{\epsilon})\; ^{F,b}\!\tilde{O}^{N}_{bare}((\alpha_s)(\mu^2/p^2)^\epsilon,\epsilon)
\label{eq27-ope}
\end{equation} 
between the renormalized and bare IR-improved matrix elements. The analoga of
(\ref{eq23-ope}) and (\ref{eq24-ope}) are then
\begin{equation}
^{F,b}\!\tilde{O}^{N}(\alpha_s,\epsilon)=\int_{-1}^{1} dx x^{N-1}\; {^{F,b}\!\tilde{O}(x,\alpha_s,\epsilon)}
\label{eq28-ope}
\end{equation} 
where 
\begin{equation}
 {^{F,b}\!\tilde{O}(x,\alpha_s,\epsilon)}=Z_F[\delta(x-1)+ x\frac{\int d^d k}{(2\pi)^d}\delta(x-\frac{kn}{pn})[\frac{\not\!n}{4kn}\tilde{T}(p,k)\not\!p]
\label{eq29-ope}
\end{equation}
and we continue to use the notation of Ref.~\cite{fur-pet1} so that 
$\tilde{T}(p,k)$ is the respective IR-improved fully connected four-point function obtained 
from the unimproved one, $T(p,k)$, by using the master formula Eq.(1) in Refs.~\cite{irdglap2} restricted to its QCD aspect, for example. This means that we get the analog of (\ref{eq25-ope}) as
\begin{equation}
Z^{-1}_{\tilde{O}}(\alpha_s,\frac{1}{\epsilon})=\int_{-1}^{1}dxx^{N-1}[\Gamma^{exp}_{qq}(x,\alpha_s,\frac{1}{\epsilon})\theta(x)-\Gamma^{exp}_{q\bar{q}}(-x,\alpha_s,\frac{1}{\epsilon})\theta(-x)]
\label{eq30-ope}
\end{equation}
where $\Gamma^{exp}_{qq},\; \Gamma^{exp}_{q\bar{q}}$ are the respective IR-improved parton densities.
We get in this way the identification of the respective IR-improved anomalous dimension as 
\begin{equation}
-\tilde\gamma^{(N)}(\alpha_s)=2\frac{\alpha_s}{2\pi}[ P^{exp}_{qq}(N,\alpha_s)+(-1)^NP^{exp}_{q\bar{q}}(N,\alpha_s)]
\label{eq31-ope}
\end{equation}
where the $ P^{exp}_{qq},\;  P^{exp}_{q\bar{q}}$ are the respective IR-improved kernels as introduced in Refs.~\cite{irdglap1,irdglap2}, where we advise that the notation
of Ref.~\cite{fur-pet1} differs from that in Refs.~\cite{irdglap1,irdglap2} by 
whether or not one includes the factor $\alpha_s/(2\pi)$ on the RHS of (\ref{eq27-ope})
in the definition of the kernels. 
This allows us to write at IR-improved one-loop level the identifications
\begin{equation}
-\tilde\gamma^{(N)}(\alpha_s)_{ij}=2\frac{\alpha_s}{2\pi}P^{exp}_{ij}(N)
\label{eq32-ope}
\end{equation}
where the labels $i,j$ span the usual values for the one-loop anomalous dimension matrix for the evolution
of the parton distributions as given in Refs.~\cite{dglap,cs, djg-fw,ghdp,hdp1} for example. This establishes in a rigorous way the
connection between the IR-improved DGLAP-CS theory in Ref.~\cite{irdglap1,irdglap2} and the OPE methods of Wilson 
as used by Refs.~\cite{djg-fw,ghdp,hdp1} in the study of deep inelastic lepton-nucleon scattering.\par
Evidently, this connection may be manifested in the analysis of other physical processes as well.
We refer the reader to Refs.~\cite{irdglap2,herwiri} wherein the new 
precision-baseline MC Herwiri1.031 which realizes the IR-improved DGLAP-CS kernels has been introduced and
compared to the Tevatron data~\cite{d0pt,galea} 
on single Z production. Its application to the various physical processes at LHC is in progress
and will appear accordingly elsewhere~\cite{elswh}, where we need to stress
that Herwiri1.031 can be applied to {\it any process} to which Herwig6.5~\cite{herwig} can be applied and that it interfaces to MC@NLO~\cite{mcatnlo} the {\it same way} that does Herwig6.5. As we have shown in Refs.~\cite{irdglap2,herwiri}, we have an improved
agreement between the IR-improved MC's shower and the Tevatron data with no need of an abnormally large
intrinsic transverse momentum parameter, PTRMS$\sim 2$GeV in the notation of Herwig~\cite{herwig}, as it is required for similar agreement with Herwig6.5~\cite{mike}.
We point-out 
that, consistent with the precociousness of Bjorken scaling, the IR-improved MC Herwiri.031 gives
us a paradigm for reaching a precision QCD MC description of the LHC data, on an event-by-event basis
with realistic hadronization from the Herwig6.5 environment, that does not involve an ad hoc hard scale
parameter, where we define ``hard'' relative to the observed precociousness of Bjorken scaling. What we have shown in the discussion above is that this paradigm has a rigorous basis in quantum field theory. 
In closing, we
thank Prof. Ignatios Antoniadis for the support and kind 
hospitality of the CERN TH Unit while part of this work was completed.\par
\par

\end{document}